# Subjective data models in bioinformatics: Do wet-lab and computational biologists comprehend data differently?

Authors:
Yo Yehudi[1], Lukas Hughes-Noehrer[1], Carole Goble[1], Caroline Jay[1]
Department of Computer Science, University of Manchester, Oxford Road, M13 9PL, United Kingdom

## Abstract

Biological science produces large amounts of data in a variety of formats, which necessitates the use of computational tools to process, integrate, analyse, and glean insights from the data. Researchers who use computational biology tools range from those who use computers primarily for communication and data lookup, to those who write complex software programs in order to analyse data or make it easier for others to do so. This research examines how people differ in how they conceptualise the same data, for which we coin the term "subjective data models".

We interviewed 22 people with biological experience and varied levels of computational experience to elicit their perceptions of the same subset of biological data entities.The results suggest that many people had fluid subjective data models that would change depending on the circumstance or tool they were using. Surprisingly, results generally did not seem to cluster around a participant's computational experience/education levels, or the lack thereof. We further found that people did not consistently map entities from an abstract data model to the same identifiers in real-world files, and found that certain data identifier formats were easier for participants to infer meaning from than others.

Real-world implications of these findings suggests that 1) software engineers should design interfaces for task performance and emulate other related popular user interfaces, rather than targeting a person's professional background; 2) when insufficient context is provided, *people may guess what data means, whether or not their guesses are correct*, emphasising the importance of providing contextual metadata when preparing data for re-use by other, to remove the need for potentially erroneous guesswork.

# Introduction

## Biology meets computers: Bioinformatics is a cross-discipline domain

Bioinformatics spans both computational and biological skillsets, as large datasets in biology necessitate computational analysis. The Alliance of Genome Resources, for example, boasts a web-portal with data curated from over one hundred and fifty thousand papers for Gene Ontology terms alone[1]. Similarly, as of July 2022, the European COVID-19 data portal[2,3] has over 12 million viral sequences and 4 million biological samples.

With such an incredible proliferation of data, **computational analysis** becomes essential. A modern biology lab is staffed not only by wet-lab biologists working with agar plates and pipettes, but also by computational biologists, bioinformaticians, data scientists and research software engineers who create tools to effectively integrate, query, analyse and visualise biological datasets, both for their own use and for use by non-computational biologists.

## Skill sets vary across bioinformatics and biology labs

When designing software interfaces for people with varying needs and skills, software engineers and user experience professionals often create "user personas" as a way of matching software features to perceived needs of the software users. These personas tend to describe specific tasks, preferences, skill sets, and goals that a user might aim to achieve whilst using the software. These personas can then be used by the software engineers to form a basis around which to design an application's features and user interfaces.

Given the range of computational skill sets in biology and bioinformatics, personas for a biological software tool might include:

1. A **wet-lab biologist** who wishes to look up information about some genes in their experiment, who generally uses a computer for emails and word processing but does not tend to write code or run computational analyses in software tools.
2. A **computational biologist or bioinformatician** with strong computational *and* biological backgrounds, who regularly writes data processing pipelines and wishes to access the same gene data and perform bulk operations on downloaded files or access the data via API.
3. A **research software engineer** who comes from a computing background but has experience creating biological software tools and wishes to access all data via API call or direct database access.

## Designing effective ways for users to comprehend data

When data are stored in a database or file, the structure of the data and relationship between different data entities is called a "data model". Considering different ways to represent data for the range of user needs described above brings us to our **research question**: Do people with differing research and experiential backgrounds in wet-lab biology and computational biology think of biological data structures in different ways? If there is a variance in thinking between skill sets, would they benefit from different background-targeted user interfaces for the same datasets?

Or, to phrase it in a different way: Does an individual's background influence data models that individuals hold inside their heads? From here on, we refer to the concept of an individual's perception of how data are shaped to be a "subjective data model".

To answer these questions, we use semi-structured interviews, background questionnaires, and user experience techniques to explore the subjective data models of 22 individuals working in biology with varying levels of computational programming skills.

## Related work

We examine selected literature related to the themes found throughout the study. This section covers mental and conceptual models in psychology and human-computer interaction; contextualised comprehension of data through diagrams and visual interfaces; and the format of meaningful identifiers.

The purpose of this study is to elicit further information around how people with biological backgrounds comprehend the "shape" of data that is relevant to their research domain, a term we call "subjective data model". This is based on previous concepts in research: Greca and Moreira[4] define "conceptual models" as an accurate and complete scientific definition of interactions within a system or event. A "mental model" is an individual's incomplete and often unstable or even incorrect personal internal representation of the same system or event.

In studies into graph comprehension, Colarusso[5] asserts that when a user is familiar with a data representation, they are able to more speedily comprehend it. Similarly, Peebles[6] recommends that when designing graphical interfaces, multiple tasks and representations of knowledge should be considered in order to aid comprehension.

Metadata is a term for "data about data" - that is, data which provides meaningful context around how data was produced, what it means, and what format it is in. Given its clarifying nature, metadata is significant to interpretation of results, but may not always be created or shared in a timely manner.

In data visualisation systems, Davies et. al[7] reflects on the tension between presenting as much underlying contextualising data as possible and allowing people to draw their own conclusions from data, vs. the risk of inadvertently allowing people to create their own (possibly incorrect) data interpretations. Overall, Davies recommends providing as much data as possible to facilitate trustworthiness of the data and decisions based on the data. This is also consistent with findings by Yehudi et al[8] around data sharing, which found that better metadata was a common wishlist item from researchers, and further that researchers had to make "educated guesses" when context was missing. Researchers found themselves unable to dispute potentially incorrect results when they were unsure of the methods and provenance used to calculate pre-processed data without access to the original raw data and associated metadata.

Multiple studies[9,10] note that even though most researchers are in favour of sharing research data, they are aware of the potential of misinterpretation without appropriate contextualisation and metadata. This may deter researchers from sharing their data publicly, as creating metadata and formatting data for re-use takes time researchers may not feel that they have available[9,11,12]. Indeed, Zhu[9] reports that 86% of participants in a UK study of academic data sharing (1,800 participants) reported that data sharing was important, but only 21% had actually shared their own data.

Finally, the file mapping portion of this study examines participants' ability to interpret alphanumeric identifiers and relate them (correctly or incorrectly) to a biological data model. McMurry et al[13] recommend that identifiers should be created as full URIs such as <https://identifiers.org/GO:0048149> where possible, as they provide context for what a given identifier might mean. They also recognise that in compact file formats (such as the files used in this study), it may be more practical to use the CURIE (Compact URI) format of the same identifier - i.e. "`GO:0048149`".

# Results

## Task A: Card Sorting task

Participants sorted the thirty-four cards onto the table. Most sorted the cards into several subgroups and topical clusters which were often loosely defined, with multiple relationships both between the clusters and between individual cards, whether or not they were in one of the clusters. We briefly describe features of the results below. Whilst the results certainly had groups and sub-types of behaviours, none of the groups seemed to cluster around backgrounds and experience. Instead, participants remarked how they would expect to interact with data based on their previous experiences with other software tools in the domain, or what task they were about to perform.

### Categories and hierarchy

Nineteen of the twenty-two participants sorted their cards or used language that either expressly asserted or indirectly implied that there were two or more levels of hierarchy within the cards provided, using phrases like "tree" and "hierarchy", "top-level", and "folder".

Thirteen participants talked about categories and clusters of cards. This largely overlapped with the group who used hierarchy in their card-sorts - only three people who used phrases like "category" did not also have hierarchy in their card arrangements.

## Most subjective data models were inherently flexible, and everything was linked

Over half of the participants - fourteen in total - indicated that they could have arranged their cards in a different way to the way that they did, or rearranged cards in response to questions about the reason for a given card being found in a certain location in the layout.

Nineteen of the participants also described links between the different trees, clusters, or piles they had made - often going hand-in hand with the above flexibility, for example (card names highlighted in **bold**):

> *"From **uniprot**, you can actually go back to the **gene**. From **gene** you can go back to **uniprot**. From both of them, you can actually go back to the **publication**. They also map to a specific **chromosome** and different **organism** so you can go with here"* (Participant 03)

> *"**Protein**s are one of the major players in **pathway**s and this can lead us to **database** which is connected with probably everything else the same as **gene**s, despite it's kind of separated. That should be arrows everywhere."* (Participant 08)

> *"So these things [list of identifiers, **BRCA, BRCA1_human, Q9H4C3_human, p53**], I couldn't identify in a simple way, [...] This is quite correlated, I could just, even sort it, and order starting with **gene** [...] I could sort it a different way, or many different ways, but it was the first idea I got."* (Participant 11)

> *"That's why I asked "how much time do we have?" I mean this is just one way of doing it. You could organise it in a different way, I'm sure other people have different entry point, they would maybe start with the example and see what they got for it - it's a **protein**, and it's transcribed from a **gene**, so you can work yourself that way through."* (Participant 14)

Context played a significant role in the way some people organised cards to define their models. Four participants indicated that they might expect to see data arranged in one way given a certain task or job role, but if their task changed they'd perhaps hope to see it organised differently, as this participant explains:

> *"It would depend very much on the context of the experiment. So over the course of my career, I started doing gene annotation for ENSEMBL, and then I might have gotten started by <u>species</u> because I was trying to do annotation for different species. When I moved into data management I was mostly doing variation work and using <u>accessions</u>,* *or using search terms in an archive. I mean I was generally given data by a collaborator, but - I sort of, I occupy an odd space, in that I'm mostly helping other people answer these questions rather than answering them myself, so <u>it would be very dependent on what question I was helping someone else answer</u>.*

### Relationships often needed qualifiers

We also found eight participants who specified properties of linked relationships, e.g. "Publication *has an* author", "Q9H4C3 *is a homologue of* BRCA1", "genes *are organised into* chromosomes", or "Genes *can be instantiated by* transcripts".

## Task B: File mapping task

Twenty people completed the file mapping task in total - the remaining two study participants either did not complete this task or it was missed from the recording.

For the purposes of the study, participants were told that there were no "correct" answers to the file mapping exercise, insofar any mapping created by a participant between a real-world identifier and a term on the provided cards would be informative. There were, however, certain cards that were more appropriate to map to specific identifiers found in the files.

For example, GO:0048149 was the identifier that had the most consistent mapping, with eighteen participants mapping it to the card "GO TERM". This identifier was indeed a Gene Ontology (GO) term that was usually identified correctly by participants. One participant (Participant 16) had never heard of GO according to their background survey but still managed to infer from the format and map it to "GO:0005515 PROTEIN BINDING". The two participants who failed to map the term to one of the two GO cards both reported that they had heard of GO but never used it.

Two other terms were largely consistent but had more variation - "`100287102`" was mapped to "`gene`", "`gene identifier`" or "`identifier`" by nineteen people, but they also mapped this identifier to a variety of additional terms, such as organism, dataset/database, accession, and name. Similarly, "`A1BG`" was mapped to "`symbol`" by fifteen people, but also "`name`" by seven people and "`gene`" by six people, with some overlap due to people choosing multiple cards per term.

Outside of these consistent terms, the other terms had much more variation in the cards mapped to them. The term with the most variation was "`FBgn0043467`".

### Participants guessed mappings, and inferred from format when they could

Identifiers which did not have CURIE-style prefixes (in the format prefix:identifier, e.g. `GO:0048149`) were sometimes mistaken for other identifier types with the same vague format. For example, two participants commented that the identifier "`1`" was strange, and three misidentified it as a chromosome when in fact, in this file, it was a gene name. A fourth participant, (Participant 5) also noted that they might have thought "`1`" was a chromosome, but they used the header information in the file to clarify that it was a gene name instead. Some quotes from participants:

> "*So this number 1, looks like it's under the column gene id. I would argue that's a stupid gene id, because it's not informative [laughter].*" (Participant 13)

> "*Yeah. Organism, 1 is gene identifier... umm, unusual to see an identifier starting with one like that.*" (Participant 4)

> "*9606? 1? no. Not without any context.*" (Participant 15, when asked directly if the first two identifiers had any meaning to them)

Similarly, the CURIE-style identifiers with prefixes allowed people to infer information from the identifiers even though they might not have been familiar with the file format, organism, or identifier in question. In total, of twenty people who completed the file mapping task, fifteen participants (three quarters of the total) used context from the prefixes ENSEMBL, GO, MIM, HGNC, or FBGN whilst explaining the way they mapped specific card terms to files.

## Task C: Data model entry points

These entry points were quite varied across participants, with twelve participants suggesting (over half) data-schema-centric entry-points such as "identifier", "accession" or "publication", five suggesting biology-centric entry points such as "gene", "disease", or "*D. melanogaster*", and two people who seemed to address elements of both information and biology when explaining their reasons for a given entry point.

**Table 1** shows the number of times a term was selected by participants. The total adds up to more than twenty-two as several participants selected multiple potential entry points, and "flymine" was cited as an entry point even though it was not in the original set of cards provided.

| Term | # of times it appeared in the entry point task |
|---|---:|
| gene | 7 |
| identifier | 5 |
| publication | 4 |
| database | 3 |
| protein | 2 |
| disease | 2 |
| organism | 2 |

| | | |
|---|---|---|
| accession | | 2 |
| pubmed ID | | 1 |
| pathway | | 1 |
| name | | 1 |
| DOI | | 1 |
| homologue | | 1 |
| flymine | | 1 |
| ensembl | | 1 |
| dataset | | 1 |
| d. melanogaster | | 1 |
| chromosome | | 1 |

Table 1: Data entry points as selected by participants

When coding the responses more broadly into "information-centric", "biology-centric", "both" or "other" we took the context of the participants' explanations into account. Of the three "other"s, one was unclear from context whether the term "gene" fell into one or both of the primary categories, one was related to electrical engineering which was the primary expertise of the participant, and perhaps most interestingly, one was from a relatively senior participant (participant 09, who had a doctoral degree in computing and biology with fifteen years experience) who suggested the entry point would be disease, "because that's where the money is".

The full table of terms suggested is shown in table 2:

| Participant ID | Entry Point(s) chosen | Category |
|---|---|---|
| Participant 01 | homologue | biology-centric |
| Participant 02 | accession, identifier, DOI, pubmed ID | information-centric |
| Participant 03 | disease, flymine, ensembl, uniprot | both |
| Participant 04 | gene | unclear |
| Participant 05 | identifier, dataset, database, publication | information-centric |
| Participant 06 | gene, identifier | both |
| Participant 07 | gene, protein, chromosome, organism | biology-centric |
| Participant 08 | gene | biology-centric |
| Participant 09 | disease | money-centric |
| Participant 10 | gene, identifier | information-centric |

| Participant 11 | protein | biology-centric |
| Participant 12 | organism, accession | information-centric |
| Participant 13 | publication | information-centric |
| Participant 14 | identifier | information-centric |
| Participant 15 | publication | information-centric |
| Participant 16 | pathway | circuits (electrical engineering) |
| Participant 17 | publication | information-centric |
| Participant 18 | database | information-centric |
| Participant 19 | d. melanogaster | biology-centric |
| Participant 20 | gene | information-centric |
| Participant 21 | gene, name | information-centric |
| Participant 22 | database | information-centric |

Table 2: Responses to "What is the entry point" question, and information-centric vs. biology-centric categories.

# Discussion

Going into this study, we expected to find specific differences between people with strongly biological (and wet-lab) backgrounds compared to people with more computational backgrounds. Whilst we did not find any evidence to support this, we did identify multiple common themes around the ways individuals perceived data models.

Three themes in particular stand out:

1. Subjective data models representing biological knowledge are both flexible and context-dependent.
2. Biological entities represented in subjective data models are highly interlinked, and sorted in hierarchical structures and categories.
3. When applying their model to real files, people infer meaning from data context, *and may guess incorrectly if no context is available.*

## Subjective data models representing biological knowledge are flexible and context dependent

The subjective data models defined by participants in the card-sort activity were usually flexible, in a way that did not appear to correlate with participant background. Participants expressed that many of the model terms, for example, could be interchangeable. This was particularly evident amongst the cluster of identifier-like terms - "name", "identifier", "accession", and "symbol".  Sixteen of the participants commented that these terms could be used interchangeably, even though five participants noted that in some communities there were specific or semantic meanings for one or more of the terms.

Another participant asserted that the quantity of data would also affect the way they might organise the data. For example, whilst the number of chromosome pairs that a human has is relatively small (twenty-three pairs), a human has tens of thousands of genes. Using genes as an entry point to the data would therefore be much harder to navigate, so they asserted they would index things differently depending on the preferred starting point they were given.

This theme repeated throughout the study from different participants, and from different angles. Our original research question asked whether individuals from biological backgrounds might have different subjective data models compared to individuals with computational backgrounds. Instead, it appears likely that subjective data models are not fixed entities, but may be flexible, and may be influenced by previous interfaces a user has experience with, or tasks that they wish to perform.

### Real-world implications of inconsistent and incorrect data mappings

This flexibility shows within a single individual - a user's individual subjective data model may change  depending upon an individual's intended task. We also observed flexibility across different individuals - that people may or may not consistently map things in the same way as their colleagues in the field would. Inconsistency has real-world implications for designing software applications. For example, in an online application designed to upload and query public data files, it would be important for data columns to be mapped consistently between different tables, if a cross-table (federated) query is to be meaningful.

Consider this example: Alice uploads a file containing fruit fly protein identifiers, selects protein as an entrypoint, and maps the protein's alphanumeric identifiers to a column called "identifier". Bob uploads a second file, with a set of papers that are related to specific fruit fly proteins. Since the papers are about proteins, he also selects protein as an entrypoint, and maps the column in his file containing the paper DOIs to "identifier". Any query returned across Alice and Bob's datasets will now be a heterogenous set of proteins and papers - almost certainly an undesirable and nonsensical result. Indeed, given the fact that an individual's subject data model may change depending on the task they are performing, it is even plausible that Alice alone could have created both these tables that resulted in the mixed protein/paper data query.

## Biological entities represented in subjective data models are highly interlinked, and sorted in hierarchical structures and categories

Whilst the subjective data models individuals described were always very flexible, they did share a few common characteristics beyond that flexibility. Participants were given no guidance as to *how* they should organise their cards when defining models, and the bulk of participants (19 out of 22) produced hierarchical category or tree-like structures, with parent terms, sub-terms, and occasionally deeper levels of nesting. We recognise that this structure could be an artefact of the data entities presented, but given that relatively few additional terms were added to the set of cards provided by participants, we believe the entities provided on cards created a logical and complete set.

Despite the fact that participants defined categories of data in their models, it was also very clear that these were not the sole links between entities.

Only one participant created categories (six piles) without any internal linkages between piles at all.

## Context is key: When applying their model to real files, people infer meaning from data context, and may guess incorrectly if no context is available.

The file-mapping exercise was designed to see whether people could map their conceptualised but abstract subjective data models to real-world data, and whether mapping the card entities to real-world data was consistent between participants. Two of the identifiers participants were asked to map to cards were "1" and "GO:0048149".

In our study, participants expressed that they found non-contextualised identifiers to be very unclear. When there was context available in the files (e.g. from file headers or identifier prefixes), participants could use that context to clarify their understanding of files. A positive example of this would be around the identifier GO:0048149, which was correctly identified as a Gene Ontology (GO) term by almost all participants, even those who hadn't used the Gene Ontology in the past.

By contrast, the identifier "1" is much more ambiguous. In the file provided, it is a genuine Homo sapiens gene identifier provided by the NCBI Entrez gene identifier system[14]. At the same time, human chromosome pairs are typically referenced by number (or X/Y in the case of sex chromosomes), so there is also a chromosome called "1". In some cases where participants did not spot the additional header context that identified that column as a gene id, they quite reasonably identified "1" as a chromosome.

This suggests that in order for data to be more usable and correctly understood, not only is contextualising metadata important, but also that the closer that context is to the data in question, the more likely it is to be correctly and effectively used. In the case of the GO term

GO:0048149, the context (GO) was directly attached to the unique part of the identifier (0048149), unlike the context for the identifier "1", where context *was* available, but further away from the unique identifier.

Perhaps more importantly, this underscores the importance of providing contextualising metadata in general when releasing data, in order to ensure that data are not inadvertently misunderstood or misused. In this study, participants repeatedly *guessed* what an identifier meant, even if they did not have sufficient context to guess correctly.

# Methods

## Ethical review

This study was approved by the University of Manchester Department of Computer Science ethics panel, approval review reference 2019-7026-11296.

## Recruitment

Participants were recruited from the EU (conference participants were invited to participate and given at least 24 hours to opt-out if they wished) and the UK, via social media and emails to institutions and individuals working in biology labs. Participants were recruited with any level of computational experience, from pure wet-lab scientists to bioinformaticians and research software engineers, but were expected to have a non-zero level of experience working or studying in biological domains, with the intent to ensure that they had some understanding / pre-existing concept of the biological processes and terms discussed in the interviews.

## Participants

We interviewed 22 people from mixed biological/computational and biological/non-computational backgrounds in semi-structured interviews. Background information was elicited via a short pseudonymised survey that asked participants about their formal biological and computational education, years of experience in biology and programming, and what biology data tools and programming tools they used. The full survey and responses are available in the supplementary data.
The majority of participants had postgraduate degrees in biology (doctoral or masters), but computational qualifications ranged from highschool / informal education only to full doctoral degrees. Years of full-time experience in biology ranged from one year to twenty-four years, with nearly half of participants (ten) having between ten to fifteen years of biology experience. Computational experience ranged from zero years to twenty-four years, and included four participants who reported that they had zero years of experience in computer programming. For the purpose of the survey, participants were instructed to treat "bioinformatics" and "computational biology" subjects as existing in both domains simultaneously.

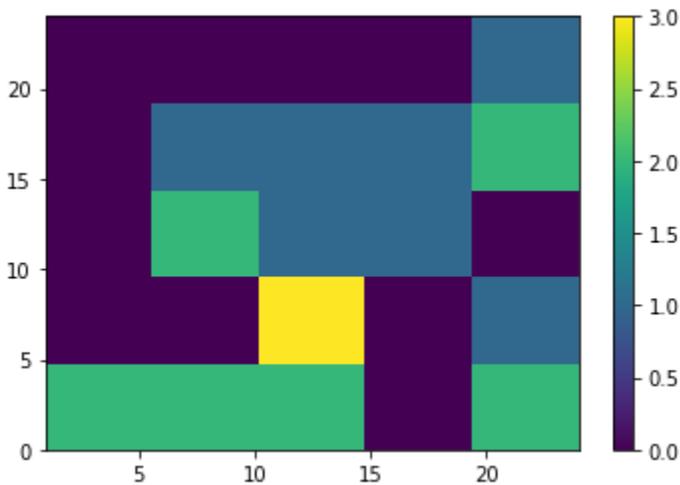

Figure 1: 2d histogram of the experience levels in biology and computer programming broken into five year buckets.

## Interview Design and Materials

Participants were provided with print-outs of three biological files and colour-coded cards: blank cards and a set of thirty-four cards which had terms written on them that directly related to biological or research data. Example terms include Gene, Publication, BRCA1, identifier, PubMed, DOI, DOI.ORG/10.1279, dataset/database, and D. melanogaster.
These cards were chosen to represent **entities** such as Gene or Publication, **attributes of these entities** such as DOI, which is an attribute of a publication or identifier, which could be an attribute of a gene, and **attribute values** such as DOI.ORG/10.1279, which is the specific DOI of a publication, or BRCA1, which is a cancer gene in Homo Sapiens (humans).

One concern while selecting these cards was whether or not the scientists we interviewed would be familiar with specific data types, values, or organisms based on their experience or research. To mitigate this, we attempted to choose a mix of "famous" identifiers where possible. For example BRCA1 is a well-known cancer-related gene, and P53 is a cancer-related protein, but we also chose some more obscure or hard to remember identifiers such as Q9H4C3_HUMAN. For the same reason, we chose sample data for cards and files across both *Homo sapiens* (human) and *Drosophila melanogaster* (fruit fly) datasets, with the hopes that we would be presenting a mix of potentially both familiar and unfamiliar data to our interviewees.

Interviewees were provided with paper, pens and had blank cards available if they wished to make notes or create any additional cards to add to the pre-provided set of cards. Interviews were video-recorded in most cases, with the exception of one recording that was audio-only at the request of the participant.

Tasks A (card sort) and B (file mapping) were alternated as starting tasks, to prevent the order of tasks from influencing overall results. Eleven participants started with the card sorting task and eleven started with the file mapping task.

## Interview Tasks

### Task A: Card sorting task

The card-sorting task was relatively free-form: participants were asked to sort the pre-provided cards on the table in a way that made sense to them, and to think aloud / explain why they were sorted in that specific way if they could. They were encouraged to add additional cards - in a different colour so they could easily be removed from the set later on - if they felt like any were missing. This task allowed participants to think of a biological data model in an abstract way, without tying it to real-world data.

### Task B: File mapping task

The file mapping task was more concrete than the card-sort, designed to mimic the process of a user importing a real data file to a database for analysis and being asked to map the columns in that file to the database's data model. This would allow us to find out whether participants consistently mapped the same files to the same properties/entities.

Participants were provided with three print-outs of the first few lines of biological files, shown in table 3. Each file had between two and six terms highlighted in blue, and participants were asked to choose zero or more cards that mapped to each of those terms, with no upper limit on cards selected per term.

Each of the three files in our file mapping task had a mix of local identifiers without a prefix, such as "`1`" or "`9606`", as well as CURIE[15]-style identifiers that had a colon-separated contextual prefix, such as "`HGNC:37102`" and "`GO:0048149`". "`FBgn0043467`" did not have the colon separation but, perhaps due to its initial alphabetic character set followed by numbers, it was treated by participants in much the same way as CURIE-style identifiers. For the purposes of this task we use the phrase "CURIE-style" identifiers to be inclusive of "`FBgn0043467`".

| File name | Highlighted terms |
|---|---|
| homo_sapiens.gff | 100287102 |
| | HGNC:37102 |
| | DDX11L1 |
| flybase_d_melanogaster.gaf | FBgn0043467 |
| | GO:0048149 |

| ncbi_homo_sapiens.gene_info | 9606 |
| --- | --- |
| | 1 |
| | A1BG |
| | MIM:138670 |
| | HGNC:HGNC:5 |
| | Ensembl:ENSG00000121410 |

Table 3: List of terms participants were asked to map card terms to.

### Task C: Entry points

After the participants had a chance to become familiar with the card set through at least one task, they were asked to identify an "entry point" into the data model - somewhere that they might start if they had a research task or experiment to perform.

## Data Preparation

After the interview data gathering stage was completed, all video and audio recordings of the interviews were transcribed and verified by a second researcher, and the original audio/video files were deleted for privacy reasons. NVivo12[16] was used for qualitative coding and thematic analysis. The second researcher also verified the common themes found whilst coding the interviews - all themes found in five or more of the interviews were verified by two separate researchers with an inter-rater reliability using Cohen's $\kappa$=0.91.

Background survey data was collected against a pseudonymised participant identifier with no personal details, using the web-based survey tool SelectSurvey. The computational data formatting and analysis pipeline was run in a python-based Jupyter notebook[17]. Data manipulations and visualisations were carried out using the Matplotlib[18,19] and Pandas libraries[20,21].

## Conclusion

This study was designed to elucidate whether individuals with varied education and experience in bioinformatics had differing perceptions of the same data structures, a concept we named "subjective data models". We anticipated that wet-lab biologists might, as a group, consistently think of data structures differently from the way computational biologists would think of the same data. To our surprise, whilst we did not find evidence that participants' subjective data models varied based on their background, we *did* find that an individual's own subjective data model tended to be highly flexible, based on the tasks they intended to perform and/or related to interfaces that they had used in the past.

Finally, when relating data models to real-world files and identifiers, participants routinely inferred information from context (when available), but took "best guesses" at other times, even when those guesses were incorrect. We conclude that providing contextualising information (metadata) is essential if researchers and data creators wish to avoid erroneous misinterpretations of their data.

# Data Availability

Data is available on the University of Manchester institutional Figshare repository, DOI [10.48420/20641017](10.48420/20641017)

# Code Availability

Analysis code is deposited on Zenodo at <https://doi.org/10.5281/zenodo.7022789> and on GitHub at <https://github.com/yochannah/subjective-data-models-analysis>. Code is shared under a permissive MIT licence.

# Author Contributions

Y.Y. designed the study, gathered and analysed the data, wrote the first draft, and approved the final version of the manuscript. L.H.N. analysed the data and approved the final version of the manuscript. C.J and C.G. supervised ideation, study design, revised the manuscript and approved the final version.

# Acknowledgements

YY would like to thank Elsa Loissel for early testing of the study design; Pete McQuilton, Aidan Budd, Emmy Tsang, Laura Clarke, Catherine Knox, and Robert Davey for assisting with recruitment; Björn Grüning for printing out updated consent forms in the middle of running the GCC2019 conference; and Rachel Lyne and Julie Sullivan for helping to select files for the file mapping task.

# Competing Interests

The author(s) declare no competing interests.